\documentclass[12pt]{iopart}
\usepackage[T1]{fontenc}
\usepackage[latin1]{inputenc}
\usepackage{subfigure}
\usepackage{floatflt}
\usepackage{graphicx}
\usepackage{setspace}
\input psfig.sty
\newcommand {\snn}	{\sqrt{s_{_{\rm NN}}}}

\newcommand {\ptass}	{p_{\perp}}
\newcommand {\pttrig}	{p_{\perp}^{\rm trig}}
\newcommand {\mpt}	{\langle p_{\perp} \rangle}

\newcommand {\deta}	{\Delta\eta}
\newcommand {\dphi}	{\Delta\phi}
\newcommand {\vflow}	{v_2}

\begin{document}

\title{Measurement of Jet Modification at RHIC}

\author{Fuqiang Wang\dag\ (for the STAR Collaboration\footnote{For the full author list and acknowledgements see Appendix ``Collaborations'' in this volume.})}
\address{\dag\ Department of Physics, Purdue University, West Lafayette, Indiana 47907, USA\\
\hspace*{0.12in}fqwang@physics.purdue.edu}

\begin{abstract}
Charged hadrons (0.15$<$$\ptass$$<$4 GeV/$c$) associated with a large $\ptass$ trigger particle (4$<$$\pttrig$$<$6 GeV/$c$) are statistically reconstructed in the large acceptance STAR TPC for p+p and Au+Au collisions at $\snn$=200 GeV. Preliminary results on transverse momentum distributions of these hadrons are presented, separately on the near and away side from the trigger particle. An increase in the multiplicity and total scalar $\ptass$ of these hadrons from p+p to central Au+Au collisions is observed. The away side distribution is found to be significantly softened in central Au+Au with respect to p+p collisions. The results are consistent with modification of jets in heavy ion collisions at RHIC.
\end{abstract}





At the Relativistic Heavy Ion Collider (RHIC), yield of large transverse momentum ($\ptass$) hadrons per binary nucleon-nucleon collision was found to be suppressed in central Au+Au relative to p+p collisions~\cite{suppression}; the strong jet-like back-to-back correlation between large $\ptass$ hadrons, as observed in p+p, was found not present in central Au+Au collisions~\cite{back2back}; d+Au collisions were found to be similar to p+p collisions~\cite{dAu}.
These results are consistent with the picture where hard-scattered partons are similarly produced in Au+Au and p+p but subsequently lose energy in central Au+Au collisions due to final state interactions~\cite{jetQuenching,hadronQuenching}. 
%
Reconstruction of the lost energy (or jets of hadrons)~\cite{theory} will therefore serve as a strong experimental confirmation of the energy loss picture. 
Moreover, by studying the amount of energy loss and how the energy is distributed, one may learn experimentally about the medium density, the underlying energy loss mechanism(s), and the degree of equilibration between the energy and the medium. 

Unlike in elementary collisions, jet reconstruction in central heavy ion collisions poses an experimental challenge due to the overwhelming background multiplicity at low $\ptass$ in these collisions. We report our attempt to statistically reconstruct jets of charged hadrons by subtracting the background in the large acceptance STAR TPC, and present preliminary results.

\section{Data Analaysis}

Details of the STAR experiment can be found in~\cite{StarNIM}. 
For this analysis, events with primary vertex within $\pm 25$ cm of the TPC center are used. 
The Au+Au events are divided into 7 centrality classes as in~\cite{back2back}.

A large $\ptass$ trigger particle is selected within the pseudo-rapidity and transverse momentum ranges of $|\eta_{\rm trig}|$$<$0.75 and 4$<$$\pttrig$$<$6~GeV/$c$. 
Charged hadrons within 0.15$<$$\ptass$$<$4 GeV/$c$ and $|\eta|$$<$1.1 
are associated with the trigger particle in $\deta=\eta-\eta_{\rm trig}$ and $\dphi=\phi-\phi_{\rm trig}$. 
%
Acceptance and tracking efficiency are corrected for the associated particles.
Background is constructed by an event-mixing technique. The elliptic flow contribution, $1+2\vflow(\pttrig)\vflow(\ptass)\cos(2\dphi)$, is included. Residual background effect is minimized by averaging the lowest data points (mostly within 0.9$<$$|\dphi|$$<$1.3) to zero. 

\begin{figure}[hbt]
\hspace*{-0.11\linewidth}
\begin{minipage}[t]{0.60\linewidth}
\hfill
\psfig{file=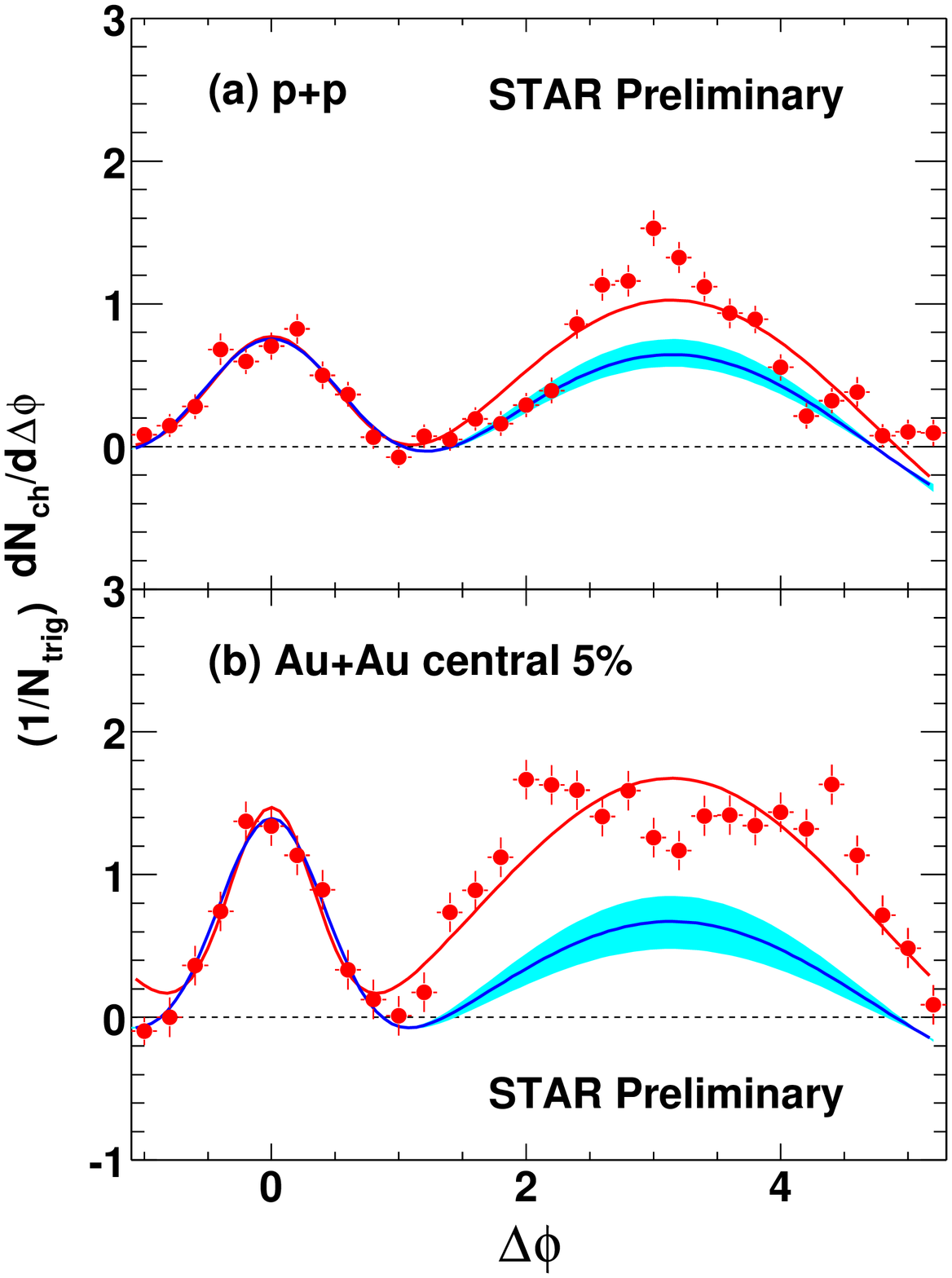,width=2.7in,bbllx=0pt,bblly=20pt,bburx=520pt,bbury=660pt}
\vspace*{-0.12in}
\caption{The per trigger particle normalized $\dphi$ distributions for p+p (a) and 5\% most central Au+Au collisions (b). See text for explanation of the curves.}
\label{fig1}
\end{minipage}
\hspace*{-0.11\linewidth}
\begin{minipage}[t]{0.61\linewidth}
\hfill
\psfig{file=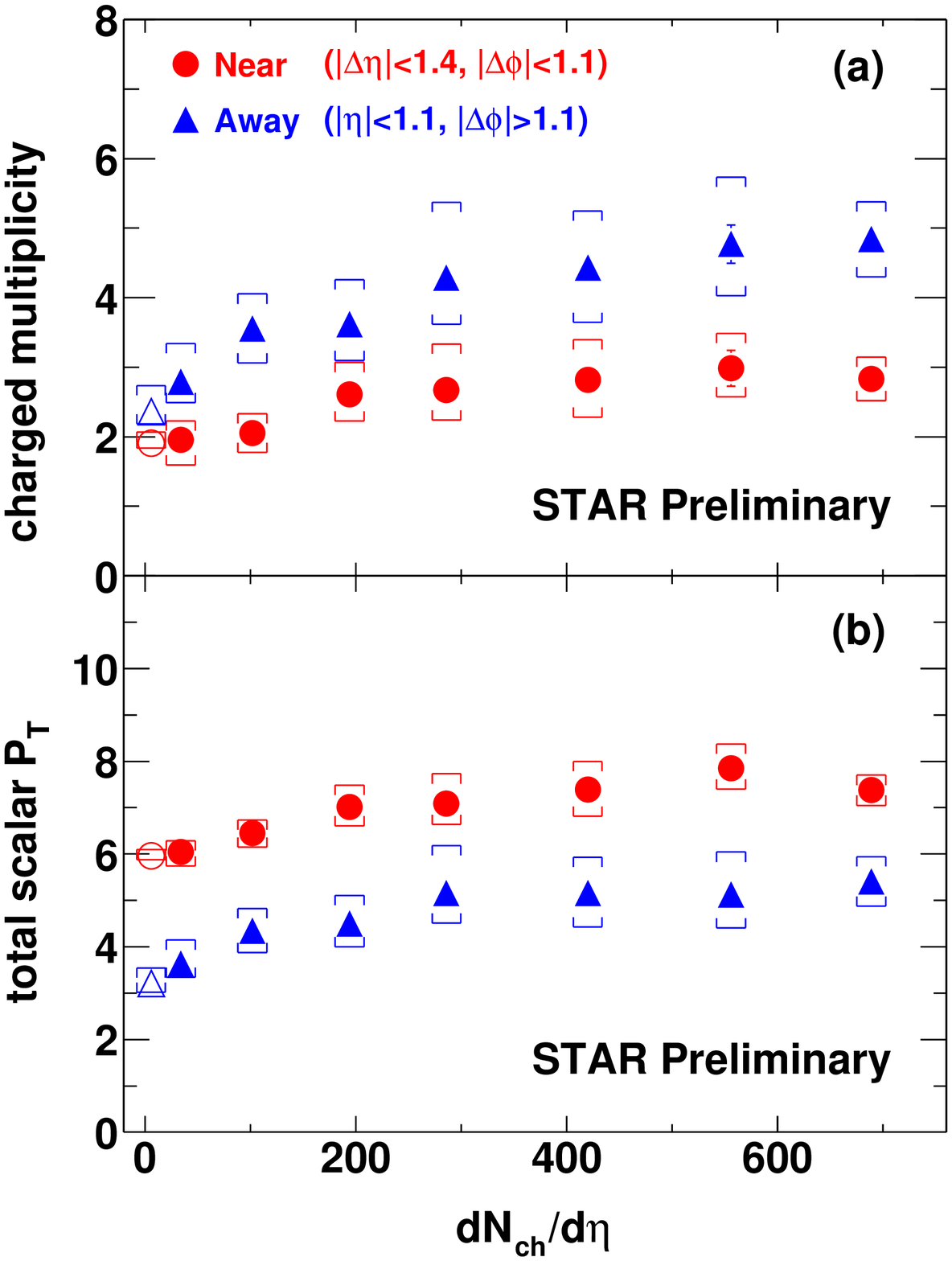,width=2.77in,bbllx=0pt,bblly=15pt,bburx=520pt,bbury=655pt}
\vspace*{-0.12in}
\caption{Charged hadron multiplicity (a) and total scalar $\ptass$ (b) on the near and away side in p+p (open symbols) and Au+Au collisions. 
Shown in caps are systematic errors.}
\label{fig2}
\end{minipage}
\end{figure}

Figure~\ref{fig1} shows the background subtracted $\dphi$ distributions.
Jet-like structures are observed in p+p and on the near side in Au+Au collisions. 
We define near side as within $|\dphi|$$<$1.1 and $|\deta|$$<$1.4 
and away side as within $|\dphi|$$>$1.1 
and $|\eta|$$<$1.1. 
We integrate the correlation signals as a measure of the jet charged hadron multiplicities. The near side signal is corrected for two-particle $\deta$ acceptance.
%
We deduce the total scalar $\ptass$ by applying a multiplicative factor of 1.5 to the associated particles to account for the neutral particles; 
we include the trigger particle $\pttrig$ for the near jet. 

\vspace*{0.1in}

\vspace*{-0.1in}
\section{Results}

Systematic uncertainties on our measurements come from three major sources:
those in (i) elliptic flow corrections, 
(ii) background normalization, 
and (iii) efficiency corrections.
The estimated systematic uncertainties are shown on the results where indicated.

Figure~\ref{fig2} shows the charged hadron multiplicity and total scalar $\ptass$ on both the near and away side in p+p and as a function of centrality in Au+Au collisions. 
The trigger particle is included in the near side results. 
We note that the away side measurements are those within the acceptance only, whereas those measured on the near side correspond to a nearly complete jet.
As seen from Fig.~\ref{fig2}, by triggering on a large $\ptass$ particle, higher energy jets are selected in central Au+Au than in p+p collisions.
The charged hadron multiplicity and total scalar $\ptass$ are found to increase from p+p to central Au+Au collisions for both the near and away side.

\begin{figure}[hbt]
\hspace*{-0.11\linewidth}
\begin{minipage}[t]{0.75\linewidth}
\hspace*{0.24\linewidth}
\psfig{file=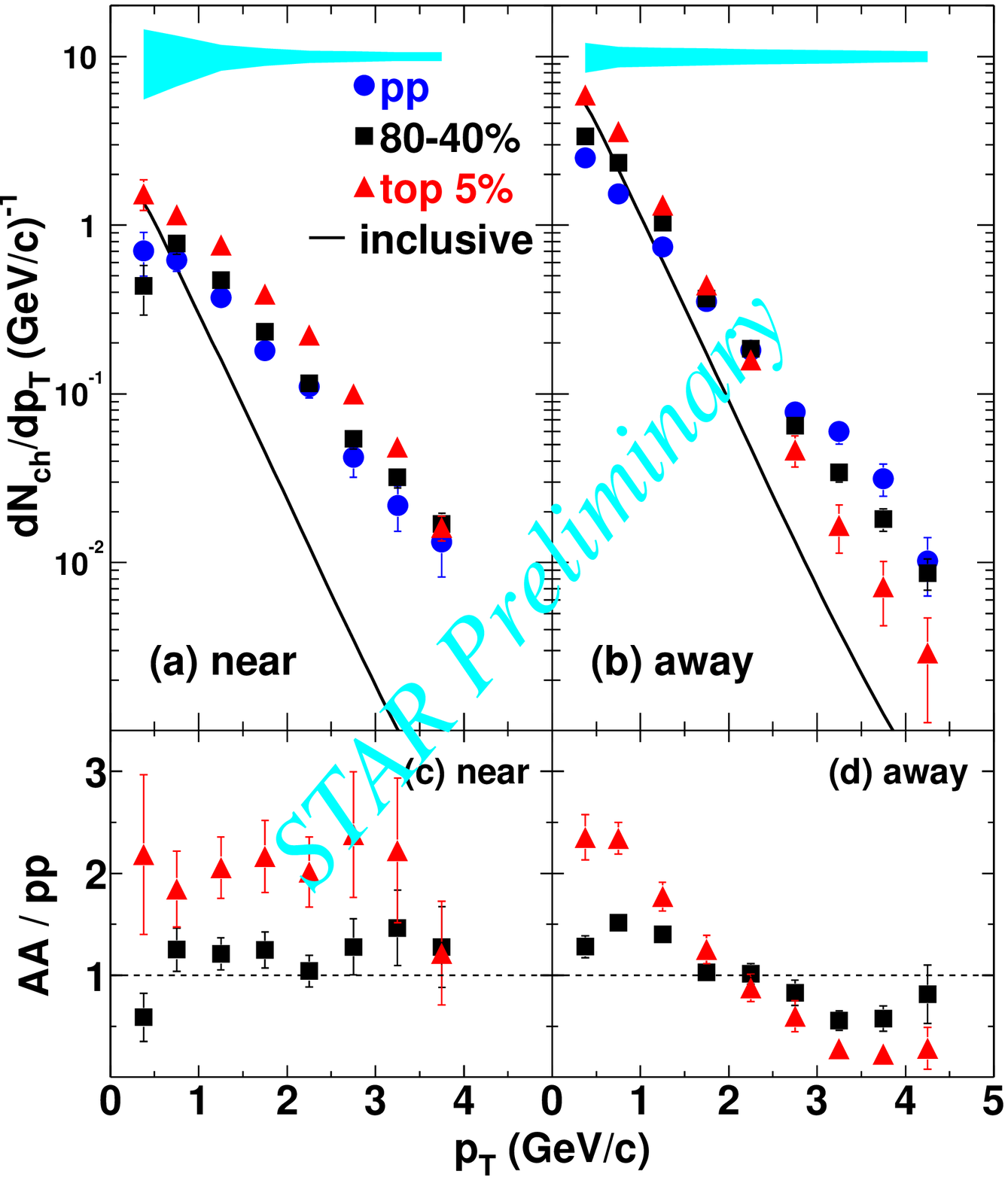,height=0.8\textwidth}
\vspace*{-0.15in}
\caption{[Upper] Near (a) and away (b) side $\ptass$ distributions of associated charged hadrons for p+p, peripheral and central Au+Au collisions.
The bands show the systematic errors for central collisions.
[Lower] Ratios of Au+Au to p+p distributions for near (c) and away (d) side. 
}
\label{fig3}
\end{minipage}
\hspace*{-0.11\linewidth}
\begin{minipage}[t]{0.45\linewidth}
\hfill
\psfig{file=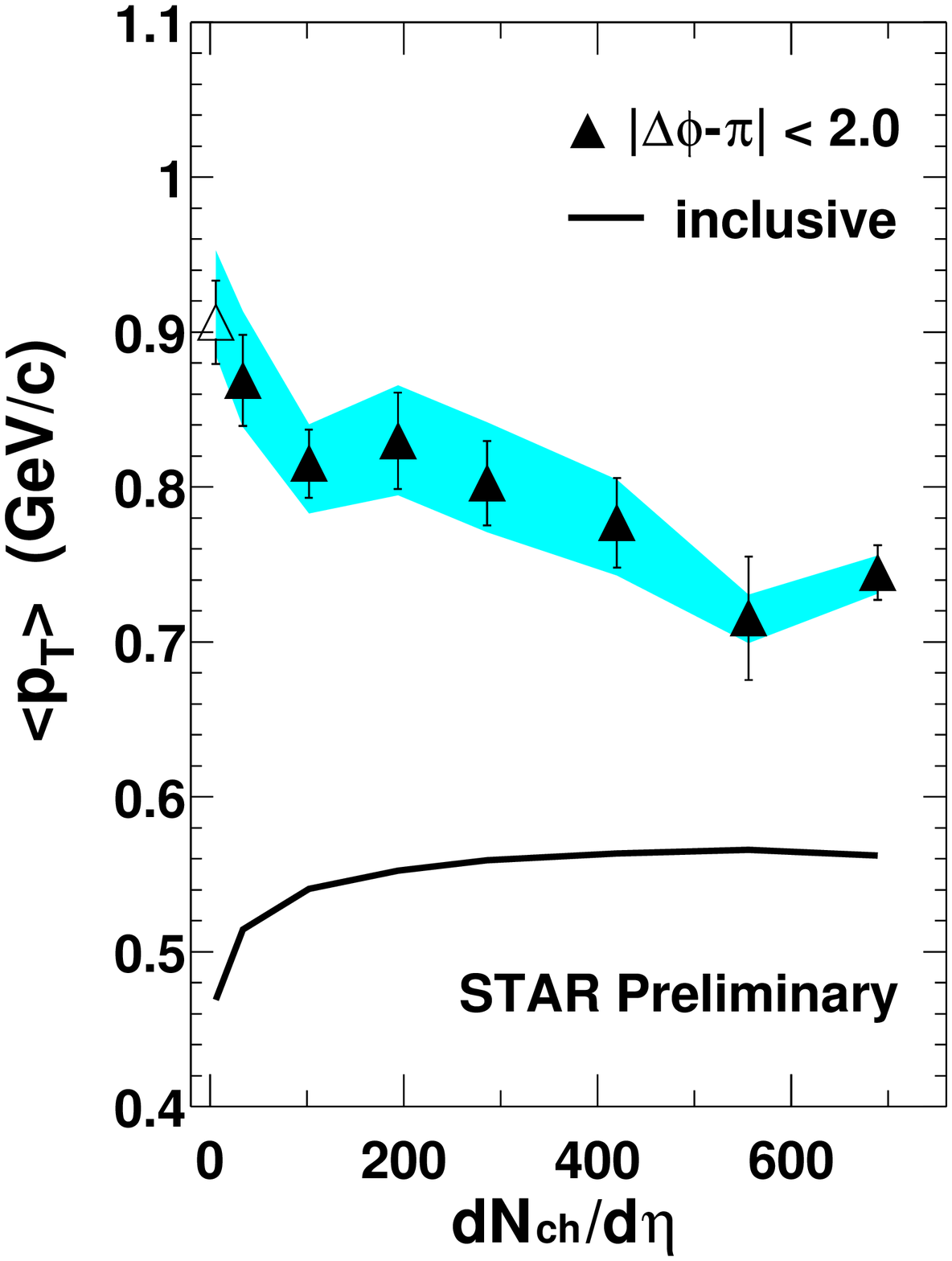,width=0.88\textwidth,bbllx=20pt,bblly=0pt,bburx=500pt,bbury=750pt,angle=0}
\vspace*{-0.33in}
\caption{Associated charged hadron $\mpt$ on the away side in p+p (open symbol) and in Au+Au collisions. Shown in the band are systematic errors.}
\label{fig4}
\end{minipage}
\end{figure}

Figure~\ref{fig3}(a-b) shows the $\ptass$ distributions of associated particles in p+p, peripheral 80-40\% and central 5\% Au+Au collisions. 
While general agreement is found between p+p and peripheral Au+Au, the central Au+Au results differ from p+p, most significantly for the away side. 
The away side in central Au+Au is depleted at large $\ptass$ and enhanced at low $\ptass$ compared to p+p collisions. 
The ratios of Au+Au to p+p distributions are depicted in Fig.~\ref{fig3}(c-d). It appears that jets 
are modified in the medium created in Au+Au collisions; the modification results in more associated particles on the near side and moves energy from high to low $\ptass$ on the away side.

\vspace*{-0.05in}

Figure~\ref{fig4} shows the centrality dependence of $\mpt$ of associated charged hadrons on the away side. 
Also shown is the $\mpt$ of inclusive hadrons that make up the medium. The away side $\mpt$ is significantly larger than that for inclusive hadrons in p+p and peripheral Au+Au collisions, and drops significantly with centrality in Au+Au collisions, approaching the medium hadron $\mpt$. This may indicate a progressive equilibration between the associated hadrons and the medium from peripheral to central collisions. 

Borghini {\it et al.}~estimated, on a statistical basis, effect on two-particle correlation due to global momentum conservation~\cite{borghini}. Such an effect would yield a correlation of $\frac{dN}{d(\dphi)}=-P_{\perp}\frac{\sum_{\rm acc.}\ptass}{\sum_{\rm all}\ptass^2}\frac{\cos(\dphi)}{\pi}$, where $P_{\perp}$ is the near side jet $\ptass$ and the summations run over all particles (denominator) and those in the acceptance only (numerator). 
An estimate of such a correlation with a constant offset is shown as the band in Fig.~\ref{fig1} (see also Ref.~\cite{Guo}).
The curve shows a free fit to the data with the sum of a Gaussian (for near side), a $\cos(\dphi)$ term (for away side), and a constant offset; it appears to describe the away side distributions well, especially for central collisions. This may suggest that the away side associated particles are reaching equilibration with the medium, in consistency with the softening of the $\ptass$ spectra shown in Fig.~\ref{fig3}(b).


\section{Conclusions}

Statistical reconstruction of charged hadron jets is performed for p+p and Au+Au collisions at RHIC by correlating charged hadrons in the $\ptass$ range of 0.15$<$$\ptass$$<$4 GeV/$c$ to large $\ptass$ particles within 4$<$$\pttrig$$<$6 GeV/$c$ and subtracting mixed-event background.
Preliminary results on charged hadron multiplicity, total scalar $\ptass$, and associated charged hadron $\ptass$ distribution are presented for both the near and away side. 
It is found, with the same $\pttrig$ leading particle in the final state, that the total scalar $\ptass$ is larger in central Au+Au than in p+p collisions, and increases gradually with centrality. 
The $\ptass$ distribution of associated charged hadrons on the away side is found to be significantly softened in central Au+Au with respect to p+p collisions, suggesting a significant equilibration of those hadrons with the medium.
The results are consistent with modification of jets in nuclear medium created at RHIC.


\end{document}